\documentstyle[12pt,fleqn]{article}
\def\beq{\begin{equation}}
\def\eeq{\end{equation}}
\def\0{\otimes}
\def\6{\langle}
\def\9{\rangle}

\begin{document}

\vspace*{15mm}

\begin{center}{\large {\bf
Construction of quantum states with bound entanglement}}\vfill

Dagmar Bru\ss $^1$ and Asher Peres$^2$\\[3mm]
{\it $^1$Institut f\"ur Theoretische Physik, Universit\"at Hannover,
D-30167 Hannover, Germany\\
$^2$Department of Physics, Technion---Israel Institute of Technology,
32000 Haifa, Israel}\vfill

{\bf Abstract}\end{center}

\begin{quote}
We present a new family of bound-entangled 
quantum states in $3\times3$ dimensions.
Their density matrix $\rho$ depends on 7 independent parameters and has 4
different non-vanishing eigenvalues.\end{quote}

\vfill\noindent PACS \ 03.67.Hk, 03.65.Bz, 89.70.+c\vfill

\newpage
Entangled quantum states have been used since the very early days of
quantum mechanics for computing the properties of atomic and molecular
systems~\cite{1927}. However, it is only in recent years that the
existence of a hierarchy of entangled density matrices became apparent,
and it is not yet fully understood.

There are classical correlations, but there is no quantum entanglement,
in {\it separable\/} density matrices that can be written as convex
sums,

\beq \rho_{m\mu,n\nu}=
  \sum_K w_K\,(\rho'_K)_{mn}\otimes(\rho''_K)_{\mu\nu}, \eeq
where the density matrices $(\rho'_K)_{mn}$ and $(\rho''_K)_{\mu\nu}$
refer to two subsystems, possibly with different dimensions, and the
coefficients $w_K$ are positive and sum up to unity. If Eq.~(1) holds,
it readily follows that the partial transpose
$\sigma_{m\mu,n\nu}=\rho_{n\mu,m\nu}$ is another separable density
matrix and in particular has no negative eigenvalue. This property gives
a very simple {\it necessary\/} condition for separability~\cite{p96}.
It also is a {\it sufficient\/} condition for systems of dimensions
$2\times2$ and $2\times3$, but not for higher dimensions~\cite{hhh96}.
The first counter\-examples for dimensions $2\times4$ and $3\times3$
contained one free parameter~\cite{pawel}.  Such states are called
``bound-entangled''~\cite{hhh98} because it is impossible to distill
from them pure singlets by means of local operations and classical
communication.

Recently, a new class of bound-entangled states was produced by means of
unextendible product bases (UPB)~\cite{upb}. In these states, the
density matrix $\rho$ depends on 6 parameters and is of rank~4 with
equal eigenvalues 0.25. If all the matrix elements are real, there are
only 4 free parameters, and $\sigma\equiv\rho$. In the complex case,
$\sigma$ and $\rho$ have similar structures, but they correspond to
different UPBs.

Here, we present a more general construction of bound-entangled states
in $3\times3$ dimensions,
depending on 7 parameters (only 5 if $\rho$ is real and $\sigma=\rho$),
with 4 different nonvanishing eigenvalues. We hope that this explicit
construction will be useful for elucidating properties of
bound-entangled states, in particular for proving (or disproving) the
conjecture that they satisfy all the Bell inequalities, and therefore
are compatible with a local hidden variable description~\cite{p99,ww}.
Other open problems are mentioned in ref.~\cite{122}.

We write $\rho$ in terms of four unnormalized eigenvectors,

\beq \rho=N\,\sum_{j=1}^4 |V_j\9\6V_j|, \eeq
where the coefficient $N=1/\sum_j\6V_j,V_j\9$ normalizes $\rho$ to unit
trace. The four eigenvalues, $N\6V_j,V_j\9$, are in general different.
Explicitly, we take

\beq \begin{array}{l}
|V_1\9=|m,0,s;\;0,n,0;\;0,0,0\9,\\
|V_2\9=|0,a,0;\;b,0,c;\;0,0,0\9,\\
|V_3\9=|n^*,0,0;\;0,-m^*,0;\;t,0,0\9,\\
|V_4\9=|0,b^*,0;\;-a^*,0,0;\;0,d,0\9,\end{array} \label{V}\eeq
where the components of $|V_j\9$ are listed in the order 00, 10, 20;
01,... It is easily seen that the 9th row and column of $\rho$ vanish.
The remaining $8\times8$ matrix is like a chessboard, with the odd-odd
components depending on $m,n,s,t$, and the even-even components on
$a,b,c,d$ (still, some of these components are zeros).

In principle, all 8 parameters on the right hand side of Eqs.~(\ref{V})
can be complex (special values of these parameters yield results
equivalent to those obtained by using UPBs). However, we still have the
freedom of choosing the overall phase of each $|V_j\9$ (this obviously
does not change $\rho$). Furthermore, we can define new phases for the
basis vectors $|e'_k\9$ and $|e''_\lambda\9$ used to describe the two
subsystems. This corresponds to rewriting $\rho$ in a different basis
without changing its chessboard structure, nor the absolute values of
the components of $|V_j\9$. This freedom can be used to make many of
these components real, but not all of them, because the two combinations
$cm/bs$ and $b^*t/n^*d$ are not affected by these changes of phase. This
can be seen as follows: the parameters in $cm/bs$ appear in
$|V_1\9=|m,0,s; ...\9$ and $|V_2\9=|...; b,0,c; ...\9$. The ratios $m/s$
and $b/c$ are affected only by changes of the relative phase of
$|e'_1\9$ and $|e'_3\9$, and $cm/bs$ is not affected at all. Likewise,
the parameters in $b^*t/n^*d$ appear in $|V_3\9=|n^*,0,0; ...; t,0,0\9$
and $|V_4\9=|0,b^*,0; ...; 0,d,0\9$. The ratios $n^*/t$ and $b^*/d$ are
affected, both in the same way, only by changes of the relative phase of
$|e''_1\9$ and $|e''_3\9$. There are no other invariants of this type,
and we can assume, without loss of generality, that $s$ and $t$ are
complex, while the 6 other parameters are real. 

We now prove that in the generic case (random parameters) $\rho$
is inseparable: as  shown in \cite{pawel}, a state $\rho$ is inseparable
if the range of $\rho$ contains no product state.  This is the case for
our $\rho$, unless the parameters are chosen in a specific way. Indeed,
assume that there is a product state such that

\beq |p,q,r\9\otimes|x,y,z\9=\sum A_j\,|V_j\9. \eeq
Since the 9th components of all the $|V_j\9$ vanish, we have $rz=0$.
Assume that $z=0$ (the same proof is valid for $r=0$, mutatis mutandis).
Then the 7th and 8th components vanish, so that $A_3=A_4=0$. We then
have $(px)(ry)=(A_1m)(A_2c)$ while $(py)(rx)=(A_2b)(A_1s)$, whence
$mc=bs$, which does not hold in general for randomly chosen parameters.

Finally, we have to verify that $\sigma$ is a positive 
matrix, so that the entanglement is bound. Namely, all the
diagonal subdeterminants of $\sigma$ have to be positive
or zero. This gives a
large number of inequalities.  Here, we shall restrict ourselves to the
study of two simple cases.

The simplest one is to assume $\sigma=\rho$. Owing to the chessboard
structure of $\rho$, this leads to three nontrivial conditions, which
can be written, with the parametrization of $|V_j\9$ in Eqs.~(\ref{V}):

\beq \begin{array}{l}
 \rho_{13}=\rho_{31}\qquad\qquad\mbox{or}\qquad\qquad ms^*=m^*s,\\
 \rho_{26}=\rho_{35}\qquad\qquad\mbox{or}\qquad\qquad ac^*=sn^*,\\
 \rho_{48}=\rho_{57}\qquad\qquad\mbox{or}\qquad\qquad ad=mt.
\end{array} \eeq
With our choice of phases, these conditions mean that $s=ac/n$ and
$t=ad/m$ are real. We thus have 6 free parameters in the vectors
$|V_j\9$. These parameters can still be scaled by an arbitrary factor
(that will be compensated by $N$), so that there are 5 independent
parameters in our construction.

Another, more general way of constructing bound entangled states is to
assume that $\sigma$ is not the same as $\rho$, but still is spanned by
two pairs of eigenvectors with a structure similar to those in
Eqs.~(\ref{V}), with new parameters that will be called  $a', b',..$.
This assumption is obviously compatible with (but not required by) the
chessboard structure of $\sigma$, which is the partial transpose of
$\rho$. It is then easily seen that the various parameters in the
eigenvectors of $\sigma$ may differ only in their phases from those in
the eigenvectors of $\rho$. We therefore write them as
$a'=ae^{i\alpha}$, and so on. This gives 8 additional arbitrary phases,
besides those of $s$ and $t$. The requirement that $\sigma$ is the
partial transpose of $\rho$ imposes 6 conditions on these phases (apart
from those on the absolute values). These are fewer conditions than
phases at our disposal, so that the parameters $s$ and $t$ can now
remain complex. Only their absolute values are restricted by

\beq |s|=ac/n\qquad\qquad\mbox{and}\qquad\qquad|t|=ad/m.\eeq
We thus have 7 independent free parameters for this case. 

A natural question is whether this construction can be generalized to
higher dimensional spaces, with a larger number of pairs of eigenvectors
$|V_j\9$, suitably structured. We have no definite answer: in such a
generalization, the number of conditions grows much faster than the
number of free parameters, and we think it unlikely that such a
generalization is possible, but we have no formal proof.

DB acknowledges support from Deutsche Forschungsgemeinschaft under
SFB 407. AP was supported by the Gerard Swope Fund and the
Fund for Encouragement of Research.

\end{document}